%% file: main.tex
\documentclass[runningheads,a4paper]{llncs}
\usepackage[T1]{fontenc}
\usepackage{graphicx}
\graphicspath{{./figures/}}
\usepackage{booktabs}
\usepackage{array}
\usepackage{multirow}
\usepackage[dvipsnames]{xcolor}
\usepackage{listings}
\definecolor{darkblue}{HTML}{1e3a5f}
\definecolor{darkred}{HTML}{7f1d1d}
\definecolor{yamlGreen}{RGB}{63,127,63}
\definecolor{yamlBlue}{RGB}{0,0,255}
\definecolor{yamlPurple}{RGB}{127,0,127}
\definecolor{yamlGray}{RGB}{100,100,100}
\lstdefinelanguage{yaml}{
  keywords={true,false,null,yes,no},
  keywordstyle=\color{yamlBlue}\bfseries,
  basicstyle=\ttfamily\scriptsize,
  sensitive=false,
  comment=[l]{\#},
  commentstyle=\color{yamlGreen}\itshape,
  stringstyle=\color{yamlPurple},
  morestring=[b]',
  morestring=[b]",
  literate={>}{{\textcolor{yamlGray}{>}}}1
           {-}{{\textcolor{yamlGray}{-}}}1
           {:}{{\textcolor{yamlGray}{:}}}1,
  escapeinside={(*}{*)}
}
\lstdefinelanguage{json}{
  keywords={true,false,null},
  keywordstyle=\color{yamlBlue}\bfseries,
  basicstyle=\ttfamily\scriptsize,
  sensitive=false,
  stringstyle=\color{yamlPurple},
  morestring=[b]",
  literate={:}{{\textcolor{yamlGray}{:}}}1
           {,}{{\textcolor{yamlGray}{,}}}1
           {\{}{{\textcolor{yamlGray}{\{}}}1
           {\}}{{\textcolor{yamlGray}{\}}}}1
           {[}{{\textcolor{yamlGray}{[}}}1
           {]}{{\textcolor{yamlGray}{]}}}1
}
\usepackage{enumitem}
\usepackage{amsmath}
\usepackage{amssymb}
\usepackage{algorithm}
\usepackage{algorithmic}
\usepackage{tikz}
\usetikzlibrary{positioning, arrows.meta, shapes.geometric, fit, calc, backgrounds}
\usepackage{fontawesome5}
\usepackage{cite}
\usepackage{placeins}
\usepackage{float}

\usepackage{hyperref}
\hypersetup{
  pdftitle={Does Runtime Topology Context Improve LLM-Generated Kubernetes Security Patches?},
  pdfauthor={Farooq Shaikh}
}

\newcommand{\kutie}{\textsc{KuTIE}}

\begin{document}

\title{Does Runtime Topology Context Improve LLM-Generated Kubernetes Security Patches?}
\titlerunning{Topology Context for LLM-Generated Kubernetes Security Patches}

\author{Farooq Shaikh\orcidID{0009-0005-3726-5373}}
\authorrunning{F. Shaikh}
\institute{Dynatrace Research, Linz, Austria \\
\email{farooq.shaikh@dynatrace.com}}

\maketitle
\raggedbottom

\input{sections/00_abstract}
\input{sections/01_introduction}
\input{sections/02_background}

\input{sections/03_architecture}
\input{sections/04_vulncare}
\input{sections/04_graph_methods}
\input{sections/05_evaluation}

\input{sections/07_conclusion}

\begin{credits}
\subsubsection{\ackname}
Claude (Anthropic) was used to assist with grammar and formatting improvements to draft prose. The author takes full responsibility for all content in this paper.
\subsubsection{\discintname}
The author has no competing interests to declare that are relevant to the content of this article.
\end{credits}

\raggedbottom
\bibliographystyle{splncs04}
\bibliography{references}

\input{sections/08_appendix}

\end{document}

%% file: sections/00_abstract.tex
\begin{abstract}
Cloud-native systems now underpin how software is built and operated at scale, and Kubernetes is
central to this ecosystem, orchestrating its containerised workloads.
Recent work suggests that large language models (LLMs) can automate cluster security remediation,
generating configuration patches and hardening recommendations from Kubernetes Security Posture
Management (KSPM) findings without human authoring.
Such systems, however, prompt the model with each finding, in isolation from the live service call
graph, assuming general hardening knowledge suffices for a correct repair.
This assumption breaks down whenever a patch must preserve a runtime service dependency invisible to
the model: an otherwise compliant fix then carries a destructive functional blast radius; crashing
downstream callers or silently severing call edges across the cluster.
Whether supplying live cluster context improves the correctness of such patches has not been
measured under controlled conditions across multiple dependency classes.

To this end, this paper introduces \kutie{} (Kubernetes Topology Intelligence Engine), which builds a
live cluster context from Istio call edges, Trivy KSPM findings, and the service-account bindings a
workload reads, and conditions LLM patch generation on it.
It is evaluated on VulnCare, a purpose-built 36-deployment, four-namespace healthcare cluster with
31~injectable findings across seven dependency classes, each labelled by topology dependence against
cluster ground truth.
Across 248~distinct trials, topology context raises topology-dependent patch correctness 
from $11.1\%$ to $78.0\%$, a gap of $\Delta = 0.669$ that holds for every model and for six of
seven classes, from credential and network-policy ($\Delta = 0.95$) to role-based access control
($\Delta = 0.31$); a topology-independent control, by contrast, exhibits no such 
effect ($\Delta = 0.0$), isolating the result from generic prompt enrichment.
Supplying the live service-call graph and the service-account bindings it exposes thus improves
remediation of topology-dependent findings well beyond what scanner-only context achieves.
\keywords{LLM \and Attack graph \and Kubernetes \and Automated security remediation \and Topology-aware patching}
\end{abstract}

%% file: sections/01_introduction.tex
\section{Introduction}

Kubernetes is an orchestration platform that runs containerised applications across a cluster of
machines. It groups an application's containers into \emph{Pods}, the smallest unit it schedules,
isolates related workloads into separate \emph{namespaces}, and exposes a single declarative
interface governing how those workloads reach one another, which privileges they hold, and what
data they may read.
This consolidation of networking, access control, and runtime privilege into one
configuration surface is both Kubernetes' operational strength and its security exposure: even a
moderate cluster spans thousands of interdependent configuration fields, and a single over-broad
setting opens an attacker foothold.
LLM-based systems can now automate the remediation of such
misconfigurations, generating corrective configurations from KSPM findings and
hardening cluster manifests without human
authoring~\cite{genkubesec2024,llmsecconfig2025,kubeguard2025,kubeintellect2025}.
These systems prompt the model with individual resource findings ranked by
severity in isolation, on the assumption that the LLM's general knowledge of
hardening guidelines suffices for producing a correct fix.
This assumption fails whenever the patch must preserve a runtime service
dependency invisible to the model.
Dropping a required Linux capability prevents a service from binding its port, so
it never starts and every downstream caller that depends on it fails;
applying a default-deny egress NetworkPolicy severs all observed call edges
from the target service; deleting an overly-broad role-based access control (RBAC)
role removes the Kubernetes API access the workload exercises at startup.
Each such fix is statically compliant yet carries a destructive \emph{functional} blast radius;
the disruption it inflicts on live cluster behaviour as opposed to the \emph{attack} blast radius
security analysis conventionally quantifies.
The same root cause, the absence of service dependency context, manifests across
capability, network-policy, RBAC, secret, credential, pod-security-policy, and
storage configuration classes, making it a structural property of the prompting
paradigm rather than a failure confined to a single misconfiguration type.

The missing element is runtime service topology: which services communicate
with which at runtime, which containers bind privileged ports to serve downstream
callers, and which compliance violations accumulate compound risk across
multi-hop service paths.
A cluster of moderate size can routinely yield hundreds of KSPM findings, yet
per-resource severity rankings conceal this path-level exposure; a LOW-severity
finding on an internet-facing service combines with a CRITICAL-severity
misconfigured backend to form a high-priority attack path that neither finding
signals independently.
Attack graph tools~\cite{mulval,kubehound} reconstruct topology from static manifest snapshots or cluster API
enumeration, producing views that diverge from the live call graph as
workloads evolve; none exposes this context for LLM generation tasks.
Runtime monitors~\cite{falco,tetragon} observe live behaviour reactively on
individual events; they do not provide structured topology context suitable for
conditioning generative model prompts.

No prior work measures, under controlled conditions, whether supplying topology
context to an LLM changes the correctness of generated security patches.
Gen\-Kube\-Sec~\cite{genkubesec2024}, LLM\-Sec\-Config~\cite{llmsecconfig2025}, and
Kube\-Intellect~\cite{kubeintellect2025} validate patches against resource schemas, and
Kube\-Guard~\cite{kubeguard2025} uses runtime logs for configuration generation.
To this end, \kutie{} collects live service-call
topology from Istio telemetry and KSPM findings from the Trivy Operator, constructs a
compliance-weighted attack graph, and conditions LLM remediation prompts on the resulting
service-call edges, bound ports, and service-account resource bindings.
The contributions are threefold.

\begin{itemize}[leftmargin=1.5em,nosep]
\item \textbf{\kutie{}.}
A system that builds a structured topology context from Istio call edges, Trivy KSPM findings, and
the service-account bindings a workload reads, ranks compliance-weighted attack paths over it,
conditions LLM remediation on the result, and gates each patch on its \emph{functional} blast radius
before application; the evaluation isolates the remediation conditioning, the ranking, annotation and
gate being described but not separately measured (\S\ref{sec:architecture},\,\S\ref{sec:graph}).

\item \textbf{The VulnCare benchmark.}
A 36-deployment, four-namespace healthcare cluster with 31~injectable findings across seven
dependency classes, each labelled topology-dependent (TD) or topology-independent (TI) against
cluster ground truth and scored by a deterministic protocol that operationalises the concept of
functional blast radius, crediting a patch only when it clears the finding while preserving the 
workload's observed call edges and readiness, released for replication (\S\ref{sec:vulncare}).

\item \textbf{A controlled methodology for isolating the topology effect.}
A factorial design labelling each finding topology-dependent or independent and scoring patches by
functional blast radius isolates topology context from generic prompt enrichment. Across 248~trials
it raises topology-dependent correctness from $11.1\%$ to $78.0\%$ ($\Delta = 0.669$), for every model
and six of seven classes (credential and network-policy $\Delta = 0.95$, RBAC $\Delta = 0.31$); the
topology-independent control shows no effect ($\Delta = 0.0$) (\S\ref{sec:rq1}--\S\ref{sec:rq4}).
\end{itemize}

%% file: sections/02_background.tex
\section{Background and Related Work}
\label{sec:background}

Work related to \kutie{} falls into three families: KSPM and static hardening,
attack-graph construction, and LLM-driven remediation. All three share a structural
limitation: findings are scoped to individual resources, edges are derived from
static configuration, and repair context centres on manifests; none conditions on
the observed runtime service-call graph.

\subsection{KSPM, Misconfiguration Detection, and Static Hardening}

Kubernetes schedules containerised workloads as \emph{Pods} across a cluster
of nodes; security configuration spans multiple resource types, including pod
\emph{securityContext} (Linux capabilities, privilege escalation), \emph{NetworkPolicy}
(inter-Pod traffic rules),
\emph{Role} and \emph{RoleBinding} (RBAC for the Kubernetes API), and \emph{Secret} and \emph{ConfigMap}
(sensitive data at rest). Each resource type carries hardening requirements codified in recommended
baselines~\cite{ciskubernetes,nsak8s}.

KSPM tools audit cluster configuration
against hardening frameworks.
Checkov~\cite{checkov} and KICS~\cite{kics} scan static manifests against policy rules;
the Trivy Operator~\cite{trivy} derives findings from live cluster state and exports them
as in-cluster Custom Resources.
On the other hand, Haque et al.~\cite{kgsecconfig2021} represent container security relationships as a
knowledge graph for posture analysis, demonstrating the value of graph structure over
flat finding lists.
KubeFence~\cite{kubefence2025} implements fine-grain API-call filtering tailored to
specific client workloads, reducing the Kubernetes API attack surface beyond what RBAC
achieves, while Bufalino et al.~\cite{insidejob2025} characterise lateral movement
arising from permissive \emph{NetworkPolicy} rules.
Runtime monitors Falco~\cite{falco} and Tetragon~\cite{tetragon} complement posture
analysis with eBPF-based syscall detection.
Across these tools the unit of analysis remains the individual resource: each finding is scored
based primarily on the resource in question, not on its position in the live service-call graph
along which its impact propagates.
\kutie{} addresses this limitation by using per-workload KSPM severity as edge weights in
an attack graph grounded in live Istio call telemetry.

\subsection{Attack Graph Construction for Cloud-Native Systems}

The formal attack graph model of Sheyner et al.~\cite{sheyner2002} represents a system
as states and attacker-executable transitions; MulVal~\cite{mulval} instantiates this
model via Datalog rules, producing high-fidelity paths at the cost of manual rule
authoring per environment.
Risk-assessment graph work extends attack graphs with countermeasures and consequences
for security assessment~\cite{unger2023risk}.
For cloud-native and infrastructure deployments, KubeHound~\cite{kubehound} models Kubernetes
API objects as typed graph nodes and enumerates attack-relevant edges from offline
manifest snapshots; and Graphene~\cite{graphene2024} analyzes infrastructure posture
through AI-generated attack graphs.
Zhang et al.~\cite{attackgplus2024} apply LLMs to construct attack knowledge graphs from
cyber threat intelligence reports, showing that language model representations recover
graph structure from unstructured text, a design principle reflected in \kutie{}'s
path annotation step.
\kutie{} constructs edges from live Istio request counters, so paths
reflect the service interactions actually executed in the cluster, and
annotates each hop against the MITRE ATT\&CK for Containers matrix~\cite{mitrecontainers}
via semantic search over the ATT\&CK Structured Threat Information Expression (STIX) bundle~\cite{stix}.

\subsection{LLM-Driven Security Automation and Remediation}

Recent work most closely related to \kutie{} applies LLMs directly to Kubernetes configuration
remediation.
GenKubeSec~\cite{genkubesec2024} detects, localises, and remediates Kubernetes misconfigurations
with 0.990 precision against rule-based baselines.
LLMSecConfig~\cite{llmsecconfig2025} augments this with retrieval-augmented generation
(RAG), reporting 94\% repair success on 1{,}000 real-world manifests.
Topology context and documentation-based RAG are orthogonal enrichment channels:
RAG retrieves static hardening guidance, whereas \kutie{} supplies observed runtime
call edges that no documentation source contains.
KubeGuard~\cite{kubeguard2025} incorporates runtime-log evidence into the hardening loop,
enabling least-privilege configuration generation that static manifest analysis alone
cannot produce.
KubeIntellect~\cite{kubeintellect2025} orchestrates modular LLM agents for end-to-end
cluster management via natural language interaction, while MetaKube~\cite{metakube2026}
adds episodic memory for Kubernetes failure diagnosis.
Beyond remediation, intent-based configuration applies LLMs to generate manifests from
natural-language operator intent: Sacco et al.~\cite{intentk8s2025} fine-tune open-source
models for this task and evaluate them with text-similarity metrics, leaving deployment-based
correctness testing and problem-specific evaluation as open challenges.
\kutie{} addresses both, scoring each patch by its effect on the running cluster rather than
by textual resemblance to a reference.
End-to-end remediation benchmarks reinforce this emphasis on functional outcome:
MicroRemed~\cite{microremed2025} evaluates whether LLM-generated playbooks actually restore
faulty microservices, though it targets runtime fault recovery rather than security
configuration under live topology.
Xu et al.~\cite{patchstudy2026} evaluate four LLM-based patching architectures,
fixed-workflow, single-agent, multi-agent, and general-purpose code agents, and show that patching
effectiveness depends on the architecture in which the LLM is embedded.
Evidence from general program repair underscores the stakes: large-scale analysis shows
LLM-generated patches frequently introduce vulnerabilities, especially for issues that omit
relevant context~\cite{aipatchsafety2025}, and adversarially crafted reports can elicit patches
that pass every test yet remain exploitable~\cite{redteamapr2025}.
\kutie{} differs by conditioning remediation prompts on the Istio-derived
call graph and by measuring, under a controlled factorial design, the
class-specific effect of topology context on patch correctness across
seven dependency classes.

%% file: sections/03_architecture.tex
\section{System Architecture}
\label{sec:architecture}

\begin{figure}[t]
\centering
\includegraphics[width=\linewidth]{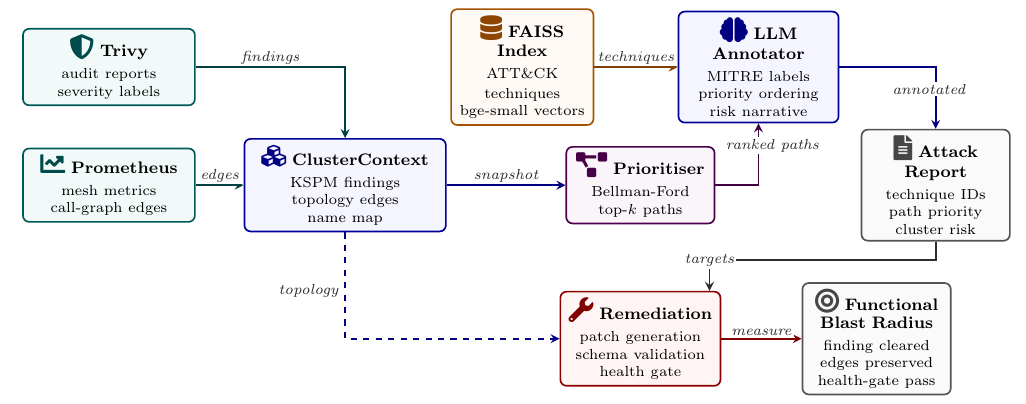}
\caption{\kutie{} pipeline: a live \texttt{ClusterContext} feeds path prioritisation, LLM
annotation, and functional-blast-radius-gated remediation.}
\label{fig:architecture}
\end{figure}

\kutie{} operates as a two-stage pipeline over a live cluster snapshot
(Figure~\ref{fig:architecture}). The first stage builds a \texttt{ClusterContext} from Istio
call telemetry and Trivy KSPM findings, ranks compliance-weighted attack paths over it, and
annotates each path against MITRE ATT\&CK to produce an analyst-facing attack report. The second
stage conditions LLM remediation on the same context and gates each generated patch on its
functional blast radius before application.

\textbf{Design rationale.}
Istio, a service mesh~\cite{istio}, is used as the topology source because its per-call counter is exported as a standard Prometheus metric,
requiring no per-service instrumentation or cluster API credentials.
Trivy Operator is used for KSPM because it publishes structured
\texttt{ConfigAuditReport} custom resources in-cluster, queryable without administrator
access, with severity classifications consistent with the Center for Internet Security Kubernetes Benchmark.

\textbf{Threat model.} The attacker is assumed to have no access to raw manifests and no ability to
enumerate the full cluster API; observable information is limited to what an external or laterally
moving workload infers from live network traffic and exposed service endpoints. \kutie{} operates
under the same constraint, driven by observed call topology and exported KSPM findings rather than
full-cluster object inspection.

\textbf{ClusterContext construction.} A \texttt{ClusterContext} object is built once per
analysis request from two data sources. KSPM findings are collected from Trivy Operator~\cite{trivy}
\texttt{ConfigAuditReport} custom resources emitted in-cluster; each finding carries a
severity classification (CRITICAL, HIGH, MEDIUM, LOW), a resource kind and workload name,
and a rule title. Call-graph edges are aggregated from Istio into directed $(src, dst)$ pairs
over a configurable scrape window (default five minutes). A workload
name resolution map normalises Kubernetes identifiers to Prometheus-observed service labels,
reconciling naming differences between the two sources.

\textbf{Technique index.} At startup, the MITRE ATT\&CK for Containers STIX bundle is loaded and
each technique description is embedded with \texttt{BAAI/bge\-small\-en\-v1.5}~\cite{bgemodel} into
a Facebook AI Similarity Search (FAISS) index.
During annotation, the concatenated KSPM rule titles are embedded with the same model
and the top-$k_t$ ($k_t = 7$) nearest techniques are retrieved, grounding the annotator in
the relevant ATT\&CK subset.

\textbf{Pipeline.} The path prioritiser constructs a compliance-weighted graph from
\texttt{ClusterContext} edges and produces a ranked path list
(\S\ref{sec:graph}). The list, per-node KSPM findings, and retrieved ATT\&CK techniques
are forwarded to the LLM annotator in a single structured prompt, whose output is an
\texttt{Annotated\-Attack\-Report}: paths ordered by comparative priority, each carrying
technique labels, confidence scores, and a cluster-level risk summary.

\textbf{Remediation loop.} Under the topology-aware condition the cluster context is
prepended to the remediation prompt as a set of labelled sections, each carrying one kind of
live-cluster fact. The objective is to clear the finding while preserving the workload's observed
service calls (minimal functional blast radius). The security context gives the service KSPM score
and its ranked attack path for situational awareness, together with the observed inbound callers,
outbound callees, and bound ports that a correct fix must keep intact. The service-account section names
the \texttt{Secret} and \texttt{ConfigMap} objects, with their keys, that the workload is
RBAC-scoped to read. Further sections report the running image digest, the cluster storage classes,
the available seccomp profiles, and the namespace cost-centre taxonomy; where applicable, two additional
context sources supply an operational constraint read from the deployment, such as the capability a port-binding
workload must retain, and the current findings of the neighbouring services on the ranked path.
The blind condition omits all of this context. Every value is read from the live cluster rather than
taken from a reference patch; the service-account binding in particular is recovered by walking
\texttt{RoleBinding} to \texttt{Role} to \texttt{resourceNames}, which exposes the exact
\texttt{Secret} a workload already reads without disclosing the correct manifest edit.
The LLM returns a strategic-merge patch validated by a server-side dry-run using \texttt{kubectl}; \kutie{} then weighs its
functional blast radius to apply the patch automatically or else flag it for human review
(\S\ref{sec:vulncare:groundtruth}). A deterministic revert and post-revert health gate ensure trial
independence.

%% file: sections/04_vulncare.tex
\section{The VulnCare Benchmark}
\label{sec:vulncare}

\subsection{Design Rationale}
\label{sec:vulncare:concept}

VulnCare is a purposefully misconfigured multi-service Kubernetes cluster,
designed as a controlled benchmark for security research that requires a ground-truth-validated
service call topology.
Healthcare is chosen as a representative domain exhibiting long service chains across external
partner boundaries (pharmacy, banking settlement, claims clearing, laboratory, and imaging
subsystems), where each link imposes a distinct security-relevant resource binding, from
privileged-port capabilities and RBAC permissions to storage-class selection and in-image
credentials.
A correct security patch on nodes in such a topology must simultaneously satisfy the KSPM
scanner and preserve every downstream call; VulnCare injects exactly this tension across the
seven dependency classes of Table~\ref{tab:taxonomy}, on a live cluster with Istio call telemetry.

The cluster runs on a three-node \texttt{kind} cluster (Kubernetes~1.32) with Cilium Container
Network Interface~1.17.4 for network policy enforcement, Istio~1.25.2 for service-mesh
telemetry, and the Trivy Operator~v0.30.1 for in-cluster KSPM scanning.
VulnCare is released as open source, with the full cluster definition, 31~injectable
findings, ground-truth scorers, the evaluation framework, and pre-computed results provided
to enable independent replication.\footnote{\url{https://github.com/dynatrace-research/vulncare}
(cluster and findings) and \url{https://github.com/dynatrace-research/kutie-artifacts}
(evaluation framework and results).}

\textbf{Why not existing clusters?}
Three open-source Kubernetes testbeds were evaluated and rejected.
Online Boutique~\cite{onlineboutique} deploys eleven microservices with no deliberate
misconfigurations; Trivy reports no topology-dependent findings, so there is no
remediation task.
Sock Shop~\cite{sockshop} has a hub-spoke topology without service-mesh telemetry,
making topology-dependent versus topology-independent discrimination impossible.
Kubernetes Goat~\cite{kubegoat} targets interactive capture-the-flag exploitation scenarios
(container escape, server-side request forgery) with single-pod challenges that do not form a
realistic multi-hop service topology.
VulnCare fills this gap: observable Istio edges, injected misconfigurations, and ground-truth
validation that naive patches break specific downstream calls.

\subsection{Service Architecture}
\label{sec:vulncare:arch}

\begin{figure}[tp]
\centering
\includegraphics[width=\linewidth]{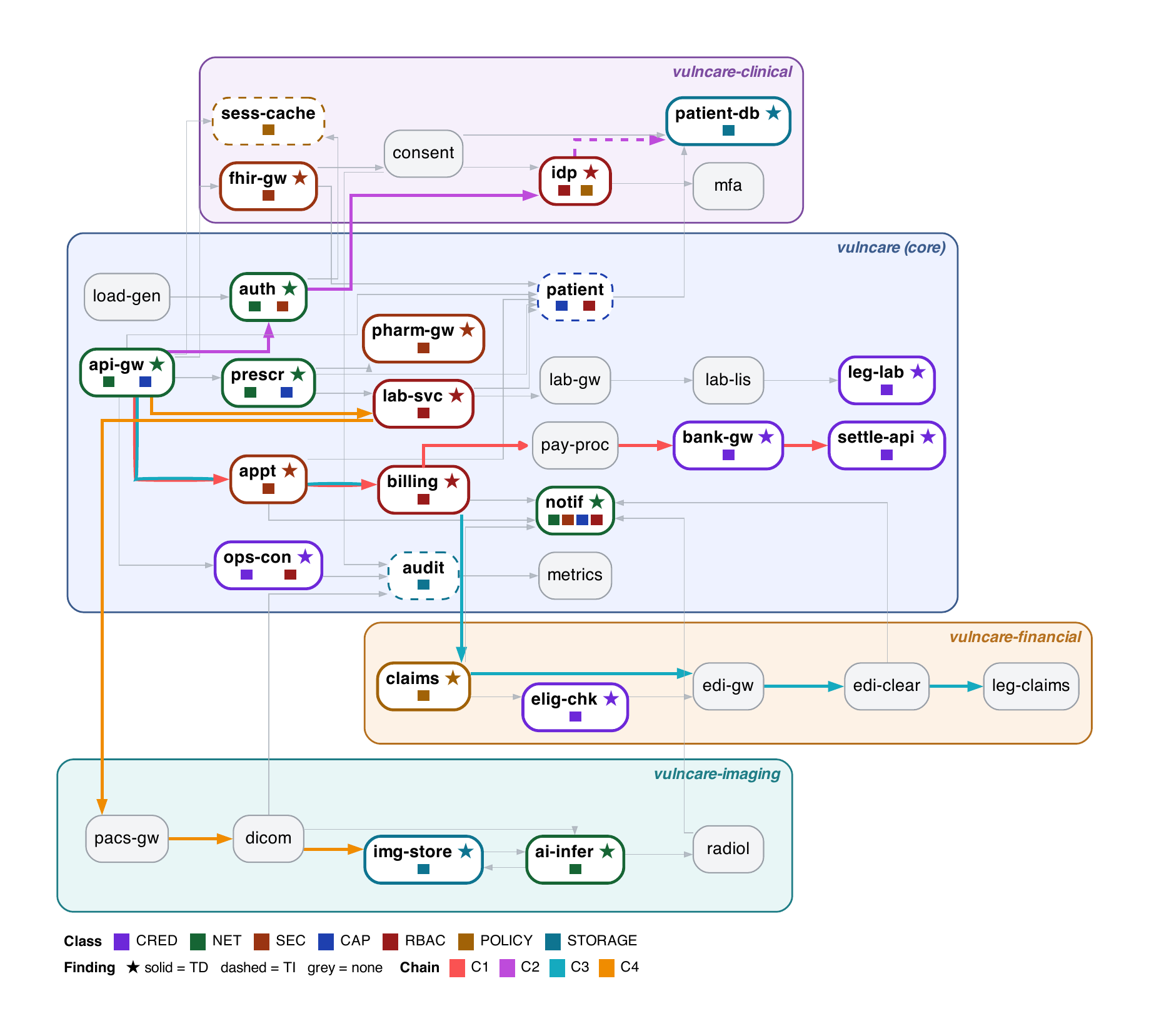}
\caption{VulnCare call graph with the per-workload misconfiguration overlay and the four
\texttt{api-gateway}-rooted exploit chains C1--C4 (\S\ref{sec:vulncare:arch}); encodings in the
legend beneath the graph.}
\label{fig:vulncare}
\end{figure}

The cluster comprises 36~deployments distributed across four Kubernetes namespaces, summarised
in Table~\ref{tab:services}.
The namespace boundary is itself a topology signal: several topology-dependent findings require
a patch whose correct value lies in a workload bound in a different namespace, so a remediation
derived from finding text alone cannot recover it.
Workloads communicate over 51~directed call edges exported from the Istio request counter; these
edges anchor the live cluster context supplied to the LLM under the aware condition (\S\ref{sec:architecture}).
Figure~\ref{fig:vulncare} renders the call graph and the misconfiguration-class overlay, with
panel boundaries identifying the four namespaces. Four \texttt{api-gateway}-rooted exploit chains
executed and validated on the live cluster are overlaid: C1 (container escape) pivots through
\texttt{bank-gateway}; C2 (RBAC escalation) reads \texttt{patient-db} credentials via the
Kubernetes API; C3 (network lateral movement) traverses six hops to
\texttt{legacy-claims-archive}; and C4 (shared persistent-volume-claim (PVC) persistence) reaches \texttt{imaging-store}.
The clinical, financial, and imaging chains each cross a namespace boundary back into the core
\texttt{vulncare} namespace.

\begin{table}[!ht]
\centering
\caption{VulnCare namespace structure. Representative workloads are those carrying injected
findings or anchoring a multi-hop chain.}
\label{tab:services}
\footnotesize
\setlength{\tabcolsep}{4pt}
\begin{tabular}{l >{\raggedright\arraybackslash}p{2.2cm} >{\raggedright\arraybackslash}p{6.0cm}}
\toprule
\textbf{Namespace} & \textbf{Function} & \textbf{Representative workloads} \\
\midrule
\texttt{vulncare}            & Core clinical, financial forwarding, lab subsystem, ops/audit
  & api-gateway, prescription-service, lab-service, lab-gateway, lab-lis-system,
    pharmacy-gateway, billing-service, payment-processor, bank-gateway,
    legacy-settlement-api, legacy-lab-archive, ops-console, audit-logger-dep \\
\texttt{vulncare-clinical}   & Identity and clinical records
  & identity-provider, mfa-service, consent-service, fhir-gateway, session-cache \\
\texttt{vulncare-financial}  & Claims and settlement clearing
  & claims-processor, eligibility-checker, edi-gateway, edi-clearinghouse,
    legacy-claims-archive \\
\texttt{vulncare-imaging}    & Medical imaging pipeline
  & pacs-gateway, dicom-processor, imaging-store, ai-inference, radiology-report \\
\bottomrule
\end{tabular}
\end{table}

\subsection{Misconfiguration Taxonomy}
\label{sec:vulncare:taxonomy}

The benchmark injects 31~findings, 25~TD and 6~TI, the latter a deliberately small negative
control, across the seven dependency classes defined in Table~\ref{tab:taxonomy}. A finding is TD when a correct, working patch requires a value
discoverable only from cluster state, and TI when the patch is fully determined by the finding text.
The Trivy finding text for all~31 is supplied to the LLM verbatim under both prompt conditions; the
per-finding taxonomy, naming the exact cluster value each TD finding requires, is given in
Appendix~\ref{sec:app:taxonomy} (Table~\ref{tab:taxonomy-full}).

\begin{table}[t]
\centering
\caption{VulnCare dependency classes: the injected misconfiguration, per-class TD/TI counts, and the
cluster-observable value a correct TD patch must recover. The full per-finding taxonomy is in
Appendix~\ref{sec:app:taxonomy}.}
\label{tab:taxonomy}
\small
\setlength{\tabcolsep}{4pt}
\begin{tabular}{l c p{3.2cm} p{4.5cm}}
\toprule
\textbf{Class} & \textbf{TD/TI} & \textbf{Injected misconfiguration} & \textbf{Cluster value a correct TD patch must recover} \\
\midrule
CAP     & 2/2 & Linux capabilities beyond need & Port a downstream caller binds \\
CRED    & 5/0 & Credentials baked into the image & RBAC-bound \texttt{Secret} name and key \\
NET     & 5/0 & Missing or over-broad egress \texttt{NetworkPolicy} & Observed caller or callee labels \\
POLICY  & 2/1 & Pod-security and image-provenance controls & None; documented patch is finding-derivable \\
RBAC    & 4/2 & API access broader than runtime needs & \texttt{resourceNames} of the bound \texttt{Secret}/\texttt{ConfigMap} \\
SEC     & 5/0 & Service credentials as plaintext env vars & RBAC-bound \texttt{Secret} name and key \\
STORAGE & 2/1 & Volume and storage-class selection & \texttt{StorageClass} or per-service PVC \\
\midrule
\textbf{Total} & \textbf{25/6} & & \\
\bottomrule
\end{tabular}
\end{table}

\subsection{Topology Differentiation Design}
\label{sec:vulncare:differentiation}

The TD/TI split is the central design instrument of the benchmark, isolating topology context from
the LLM's general hardening knowledge. Every TD finding admits a KSPM-compliant but topology-neutral
patch that passes the static rule yet breaks a downstream dependency or fails the exact-match
criterion, whereas a TI finding is fixable from its text alone; the TI entries thus form a
within-class control separating a genuine topology effect from generic prompt enrichment.

Each class value (Table~\ref{tab:taxonomy}) is cluster-discoverable state, the same introspection a
human operator performs, not the scored answer: a patch earns $\alpha = 1$ only when the model
itself selects and transcribes the correct value. Under blind prompting the model cannot access
these values, the structural reason the blind condition fails on TD findings irrespective of model
capability.

\subsection{Ground-Truth Validation}
\label{sec:vulncare:groundtruth}

Ground truth is derived from the live cluster: for each finding the dependency-preserving value was
established from observed call edges, service-account bindings, and the cluster objects a workload
reads. Scoring is deterministic and static. A per-class scorer inspects the submitted patch and
assigns $\alpha = 1$ only when it clears the KSPM finding and carries the value ground truth records
as dependency-preserving, and $\alpha = 0$ to a structurally valid but topology-neutral patch.
Exact-value recovery is the stricter predicate and serves as a mechanised proxy for functional blast
radius rather than a per-trial re-observation of the cluster; it can therefore score a functionally
healthy but over-permissive patch as incorrect, which depresses the aware arm.
The breakages are class-specific. Removing the required capability on a CAP-TD finding leaves the
workload unable to bind its port, so it never becomes ready and its callers lose the dependency; a
default-deny \texttt{NetworkPolicy} on a NET-TD finding cuts the observed call edges, and the
affected requests start to fail. For CRED-TD, SEC-TD, and RBAC-TD findings the discriminating value
is the exact \texttt{Secret} or role identifier (and, for credentials, its key): naming the wrong
object prevents the container from starting even though the scanner finding clears. STORAGE-TD needs
a cluster-specific \texttt{StorageClass} the finding text lacks; POLICY-TD is the exception, its
documented patch being finding-derivable. The six TI findings check the converse: topology-neutral and
topology-aware fixes produce identical outcomes, confirming they carry no topology-discriminating
signal.

%% file: sections/04_graph_methods.tex
\section{Graph-Based Attack Path Analysis}
\label{sec:graph}

\subsection{KSPM-Grounded Path Prioritisation on a Compliance-Weighted Service Graph}

A weighted directed graph $G_S = (V_S, E_S)$ is built from live Istio call edges;
edge weights encode KSPM severity at the source node:

\begin{equation}
w(u,v) = \begin{cases}
  0.4 & \text{if } \sigma(u) \geq 10 \quad \text{(CRITICAL finding present)} \\
  0.7 & \text{if } \sigma(u) \geq 5  \quad \text{(HIGH finding present)} \\
  1.0 & \text{if } \sigma(u) \geq 2  \quad \text{(MEDIUM finding present)} \\
  1.5 & \text{otherwise}
\end{cases}
\end{equation}

where $\sigma(u) = \sum_{f \in \text{KSPM}(u)} w_f \cdot c_f$ sums severity-weighted
finding counts, with $w_f \in \{10,5,2,1\}$ a monotone ordinal mapping of the four severity levels
and $c_f$ the per-rule finding count.

\textbf{Reward function.} A per-node reward balances local severity against neighbourhood severity:

\begin{equation}
r(v) = 0.6 \cdot \sigma(v) + 0.4 \cdot \bar{\sigma}(v)
\end{equation}

where $\bar{\sigma}(v)=\sum_{t\in\mathcal{N}^+(v)}\sigma(t)$ sums out-neighbour severities;
$w_{\mathit{eff}}(u,v)=w(u,v)\cdot(1-\tilde{r}(v))$, $\tilde{r}(v)=r(v)/r_{\max}\in[0,1]$.

\textbf{Implemented prioritisation heuristic.} Bellman-Ford relaxation is used in place of Dijkstra's algorithm to keep the prioritiser correct should edge weights later admit negative, reward-dominated values; on the present non-negative weights the two coincide, and since $w_{\mathit{eff}}(u,v) \geq 0$, no negative-weight cycles arise and convergence is guaranteed within $|V_S|-1$ passes.

\textbf{Path scoring.} The final combined score for a path terminating at node $v$ is:

\begin{equation}
\text{score}(v) = \bar{r}(v) - \frac{d(v)}{d_{\max}}
\end{equation}

where $\bar{r}(v)$ is the mean normalised reward and $d(v)$ the accumulated distance normalised by $d_{\max}$. Paths are ranked in descending order; the top-$k_p$ ($k_p = 10$) are forwarded to the LLM annotator.

\textbf{Scope.} Because $G_S$ contains only service-level edges, pod-level attack primitives
including RBAC privilege escalation, hostPath volume abuse, and container escape are not
modelled. The prioritiser answers \emph{where} an attacker can reach within the observed service graph;
\emph{how} each hop is realised is the role of the LLM annotator.

\subsection{LLM Annotator: MITRE Technique Assignment and Priority Ordering}

The LLM annotator receives the full ranked path list produced by the prioritiser and performs two
tasks in a single structured inference call: path-level MITRE ATT\&CK annotation grounded in
KSPM evidence, and comparative priority ordering across all candidate paths.

\textbf{Input construction.} The prompt supplies the top-$k_t$ ATT\&CK techniques retrieved
from the FAISS index, the full ranked path list with per-path scores, and the per-node KSPM
findings for every node appearing in any path.

\textbf{Hop annotation.} For each hop $(u \to v)$ the annotator emits the content of the
analyst-facing attack report: a candidate MITRE ATT\&CK technique, the KSPM rule titles at $u$ that
enable it, and a confidence derived from the maximum KSPM severity at $u$. Grounded in the retrieved
ATT\&CK definitions and the observed KSPM evidence, the suggested techniques stay tied to the
findings that motivate them, and the per-hop confidences combine into a path-level score as their
geometric mean.

These annotations characterise attack paths, the routes along which an adversary could traverse the
cluster, while their compliance-weighted ranking surfaces the multi-hop exposures that most warrant a
security analyst's attention. To establish that this ranking reflects genuinely exploitable routes
rather than its highest-scoring entries alone, five chains spanning the ranking from a top-ranked
path to the deepest and lowest-ranked were executed end to end on the live cluster, the four in
Figure~\ref{fig:vulncare} with one further chain, and each proved exploitable
(\S\ref{sec:vulncare:arch}).

Separately from the prioritiser's KSPM cost rank, the annotator assigns each path an
attacker-priority rank, from 1~(highest) to $k$~(lowest) across all candidate paths, by the severity
and specificity of its KSPM evidence, reachability from an unauthenticated entry point, terminal-node
impact, and path feasibility.

\textbf{Output.} The annotator returns a structured \texttt{Annotated\-Attack\-Report}:
a list of \texttt{Annotated\-Path} records sorted by comparative priority. Each record carries
the path annotations, a 2--4 sentence exploitation narrative for the end-to-end path, the
priority rationale, and a cluster-level risk summary paragraph. This output is the analyst-facing
report.

\subsection{Path Prioritisation on VulnCare}
\label{sec:graph:results}

Run over the benchmark cluster (\S\ref{sec:vulncare}), the
prioritiser enumerates 33~ranked paths from the \texttt{api-gateway}\linebreak ingress, of depth one to six
hops $\{1{:}8,\,2{:}9,\,3{:}6,\,4{:}6,\,5{:}3,\,6{:}1\}$. The top-ranked path is not the shortest:
the three-hop lab chain from \texttt{api-gateway} through \texttt{lab-service} and
\texttt{lab-gateway} to \texttt{lab-lis-system}
scores $0.74$ and outranks all eight directly reachable services, because accumulated KSPM severity
along the lab subsystem outweighs proximity to the entry node. Severity-weighted ranking therefore
surfaces finding-dense chains that hop-count ordering buries beneath shallow neighbours.

%% file: sections/05_evaluation.tex
\section{Evaluation}
\label{sec:eval}

The evaluation addresses four research questions:
(RQ1)~does topology context improve patch correctness on topology-dependent misconfigurations;
(RQ2)~which dependency classes benefit most, and which least;
(RQ3)~on topology-independent misconfigurations, does topology context change correctness, as a
control for generic prompt enrichment; and
(RQ4)~what failure modes account for incorrect patches under each condition.

\subsection{Measurement Methodology}
\label{sec:setup}

\textbf{Corpus and design.}
The benchmark corpus is the 31~VulnCare findings (25~TD, 6~TI; \S\ref{sec:vulncare}).
The experiment crosses 31~findings $\times$ 4~models $\times$ 2~conditions for 248~trials.
Under the \emph{aware} condition the prompt includes the cluster context
(\S\ref{sec:architecture}): adjacent Istio call edges, bound ports, and the service-account
binding that names the \texttt{Secret}, \texttt{ConfigMap}, or role the workload already reads.
Under \emph{blind}, only the Trivy finding text and the current deployment spec are supplied.

\textbf{Models.}
Four LLMs are evaluated: two proprietary frontier models
(Claude Sonnet~4.6, Haiku~4.5) and two open-weight alternatives (Llama~4 Maverick,
Mistral~Large). All calls are single-turn at temperature~0.

\textbf{Scoring and ground truth.}
Ground truth is derived from the live cluster (\S\ref{sec:vulncare:groundtruth}); a deterministic
per-class scorer then applies it across every trial by static inspection of the submitted patch,
as binary correctness $\alpha \in \{0,1\}$. A TD patch is
credited with $\alpha = 1$ only where it both
clears the KSPM finding and recovers the cluster-specific value the class requires, whereas a
structurally valid yet topology-neutral patch receives $0$. One blind trial errored at generation
(\texttt{STORAGE-TD-2}~$\times$~Llama, malformed JSON) and is excluded from its denominator,
leaving 99~blind TD trials.

\textbf{False-negative bound.}
Within the 25~TD findings, three blind trials score $\alpha = 1$ for reasons unrelated to topology:
 \texttt{RBAC-TD-1}~$\times$~Sonnet (the injected manifest still
carried a \texttt{resourceNames} hint), \texttt{CAP-TD-2}~$\times$~Llama (the finding text named the
required capability), and \texttt{SEC-TD-2}~$\times$~Llama (the \texttt{Secret} name coincides with
the env-var name). These findings set a conservative upper bound on blind correctness.

\subsection{RQ1: Topology Context and TD Correctness}
\label{sec:rq1}

\textbf{Result.}
Topology context raises TD patch correctness from $11.1\%$ ($11/99$) under blind prompting to
$78.0\%$ ($78/100$) under aware prompting, a gap of $\Delta = 0.669$. Across the 99~paired
finding--model trials, 70 of the 73~discordant pairs favour the aware condition and 3 the blind
(McNemar's exact test, $p \approx 1\times10^{-17}$). The topology-aware advantage
on TD findings is the dominant effect in the study and holds across every model
(Table~\ref{tab:per_model}) and six of seven dependency classes (Table~\ref{tab:per_class}).

\textbf{Robustness across models.}
The per-model gap ranges from $0.59$ (Llama~4 Maverick) to $0.76$ (Claude Sonnet)
(Table~\ref{tab:per_model}); the effect is a property of the prompting condition, not of one family.

\begin{table}[htbp]
\centering
\caption{Per-model TD patch correctness ($\alpha$-rate over 25~TD findings) under aware and blind
conditions.}
\label{tab:per_model}
\small
\setlength{\tabcolsep}{6pt}
\begin{tabular}{lccc}
\toprule
\textbf{Model} & \textbf{Aware} & \textbf{Blind} & $\boldsymbol{\Delta}$ \\
\midrule
Claude Sonnet~4.6 & 22/25\ \ (0.88) & 3/25\ \ (0.12) & $0.76$ \\
Claude Haiku~4.5  & 19/25\ \ (0.76) & 2/25\ \ (0.08) & $0.68$ \\
Llama~4 Maverick  & 19/25\ \ (0.76) & 4/24\ \ (0.17) & $0.59$ \\
Mistral~Large     & 18/25\ \ (0.72) & 2/25\ \ (0.08) & $0.64$ \\
\midrule
\textbf{Overall}  & \textbf{78/100\,(0.78)} & \textbf{11/99\,(0.11)} & $\mathbf{0.67}$ \\
\bottomrule
\end{tabular}
\end{table}

\subsection{RQ2: Per-Class Breakdown}
\label{sec:rq2}

Table~\ref{tab:per_class} reports TD correctness by class. Six of the seven classes show a positive
gap varying more than threefold; POLICY alone is negative, and the interpretation below treats it
separately as evidence of context-induced blast radius.

\begin{table}[htbp]
\centering
\caption{Per-class TD patch correctness under aware and blind conditions.}
\label{tab:per_class}
\small
\setlength{\tabcolsep}{6pt}
\begin{tabular}{lcccc}
\toprule
\textbf{Class} & $n$ & \textbf{Aware} & \textbf{Blind} & $\boldsymbol{\Delta}$ \\
\midrule
CRED    & 5 & 19/20\ (0.95) & 0/20\ (0.00) & $0.95$ \\
NET     & 5 & 19/20\ (0.95) & 0/20\ (0.00) & $0.95$ \\
SEC     & 5 & 17/20\ (0.85) & 1/20\ (0.05) & $0.80$ \\
CAP     & 2 & \ 7/8\ \ (0.88) & 1/8\ \ (0.12) & $0.75$ \\
STORAGE & 2 & \ 5/8\ \ (0.62) & 0/7\ \ (0.00) & $0.625$ \\
POLICY  & 2 & \ 5/8\ \ (0.62) & 8/8\ \ (1.00) & $-0.375$ \\
RBAC    & 4 & \ 6/16\ (0.38) & 1/16\ (0.06) & $0.31$ \\
\bottomrule
\multicolumn{5}{p{0.78\linewidth}}{\scriptsize $n$: TD findings in the class. Aware/Blind:
correct trials over non-errored trials across four models.} \\
\end{tabular}
\end{table}

\begin{figure}[tbp]
\centering
\includegraphics[width=\linewidth]{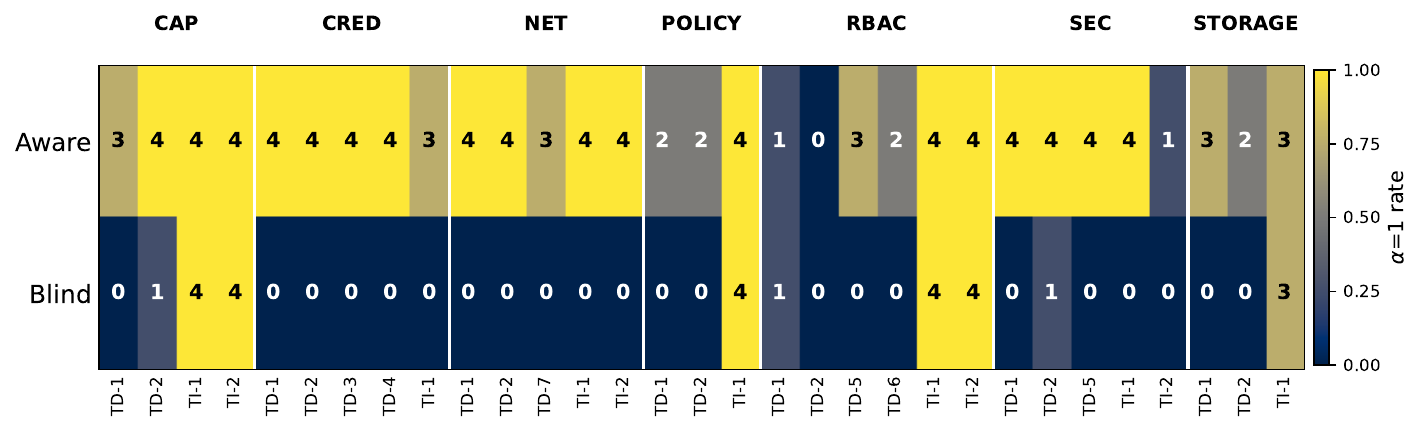}
\caption{Per-finding patch correctness. Tile colour is the $\alpha{=}1$ rate over the four models
(cividis, lighter is higher); the printed digit is how many of the four produced a correct patch.
Columns are the 31~findings grouped by class; rows are aware vs blind prompting.}
\label{fig:heatmap}
\end{figure}

\textbf{Interpretation.}
The credential and network classes benefit most. For CRED and SEC the correct patch must reference
a pre-existing \texttt{Secret} by name and key, recoverable only from the service-account binding;
for NET it must name the observed caller or callee. Under blind prompting no model produces any
correct NET or CRED patch (both $0/20$), and only a single SEC patch succeeds, by name coincidence.
RBAC benefits least among positive classes ($\Delta = 0.31$): even with the binding supplied, models
emit a \texttt{Role} scoped to the right resource type but omitting the precise
\texttt{resourceNames} entry. POLICY inverts ($\Delta = -0.375$): the supplied image digest induced
three of four models to repoint the \texttt{claims-processor} image, severing the
\texttt{eligibility-checker} and \texttt{edi-gateway} edges the correct annotation-only patch keeps.
Figure~\ref{fig:heatmap} shows the per-finding pattern across both conditions.

\subsection{RQ3: Topology-Independent Control}
\label{sec:rq3}

The six topology-independent findings act as a within-study negative control: their correct patch
is fully determined by the finding text (drop a named capability, delete a named \texttt{ClusterRole},
add a size limit, add a chargeback annotation), so topology context should not change the outcome.
The results clearly show it does not. Aware and blind correctness are identical at
$23/24$ ($95.8\%$) each, $\Delta = 0.0$;
the single miss (\texttt{STORAGE-TI-1}~$\times$~one model) occurs under both conditions. The contrast
with the TD gap of $0.669$ confirms that the topology-aware advantage is specific to findings whose
correct remediation requires a cluster-observed value, not a generic effect of a longer prompt.

\subsection{RQ4: Failure Modes}
\label{sec:rq4}

Under blind prompting, $88$ of the $99$ TD trials fail with one shared structure: the model emits a
standard hardening patch that clears the scanner finding while omitting the cluster-specific value.
The recurring forms are an invented \texttt{Secret} name or key (CRED, SEC), a default-deny policy
that severs the observed edge (NET), a \texttt{drop:~ALL} set on a port-binding service (CAP), and a
default \texttt{StorageClass} or dangling claim reference (STORAGE). These are not malformed
patches but plausible, scanner-compliant fixes with a destructive \emph{functional blast radius}, failing
only because the runtime dependency is invisible to the model.

Under aware prompting, the $22$ residual failures concentrate in RBAC ($10$ of $16$), POLICY ($3$ of
$8$), and STORAGE ($3$ of $8$). The common factor is transcription rather than reasoning: the binding
is present, but the model narrows or paraphrases the supplied value rather than copying it exactly,
in the per-class forms detailed in \S\ref{sec:rq2}. The credential and network classes, where the
value is a single name supplied verbatim, show almost no aware failures. The remaining gap is
therefore a limit on transcribing a supplied value, not evidence that the value was unavailable.
Prompting alone does not close it: slot-filling the patch template, or validating emitted values
against the supplied context, targets transcription where the dry-run in place cannot.

\FloatBarrier

%% file: sections/07_conclusion.tex
\section{Conclusion}
\label{sec:conclusion}

Conditioning Kubernetes security remediation on observed service-call topology and the
service-account bindings a workload exercises, rather than on the scanner finding alone, changes what
an LLM can fix. Across 248~controlled trials on VulnCare, topology context raised topology-dependent
patch correctness from $11.1\%$ to $78.0\%$ ($\Delta = 0.669$), consistently across every model and
six of seven dependency classes, the gain scaling with how much of a correct fix reduces to a single
cluster-observed value, from credential and network policy down to RBAC
(\S\ref{sec:rq1}--\S\ref{sec:rq2}).

This effect is specific to findings that require such a value. A topology-independent control showed
no aware--blind difference, ruling out generic prompt enrichment, while the characteristic blind
failure was uniform, a scanner-compliant patch that omits the runtime-required value and so carries a
destructive functional blast radius, structural to finding-isolated prompting rather than an artefact
of any one model (\S\ref{sec:rq3}--\S\ref{sec:rq4}).

\kutie{} applies wherever a deployment exposes service-account bindings and an Istio-class layer-7
telemetry source; recovering the bindings needs read access to \texttt{RoleBinding} and
\texttt{Role}, so mesh telemetry alone does not suffice. Three caveats qualify the result. The call
graph captures service-to-service edges rather than pod-level escalation primitives, so recovered
paths express lateral reachability rather than per-hop exploit mechanics. Edges are aggregated over
a finite scrape window, so rarely exercised paths may be absent, and a missing edge yields the
destructive patch the approach exists to prevent. The effect was measured on a single healthcare
cluster and four models, and its magnitude depends on the 25/6 TD/TI composition chosen by
construction; how often TD findings arise in production is not established here. All experiments ran
on isolated \texttt{kind} clusters, with no production systems or personal data involved.

%% file: sections/08_appendix.tex
\appendix
\clearpage
\section{Full Misconfiguration Taxonomy}
\label{sec:app:taxonomy}

Identifiers are non-contiguous: findings dropped as topology-neutral during construction, before
the reported trials, keep their numbering.

\begin{table}[H]
\centering
\caption{VulnCare misconfiguration taxonomy: 31~findings across seven classes (25~TD, 6~TI).}
\label{tab:taxonomy-full}
\scriptsize
\renewcommand{\arraystretch}{0.92}
\setlength{\tabcolsep}{2pt}
\begin{tabular}{l l l c p{4.7cm}}
\toprule
\textbf{ID} & \textbf{Cl.} & \textbf{Service} & \textbf{Dep.} & \textbf{Topology signal} \\
\midrule
CAP-TD-1     & CAP & api-gateway          & TD & Caller graph: retain only downstream-used port capability \\
CAP-TD-2     & CAP & prescription-service & TD & Callee ports determine which capabilities to keep \\
CAP-TI-1     & CAP & patient-service      & TI & Drop named capability (in finding text) \\
CAP-TI-2     & CAP & notification-service & TI & Drop named capability (in finding text) \\
\midrule
CRED-TD-1    & CRED & bank-gateway          & TD & \texttt{core-banking-secret}; depth-5 financial chain \\
CRED-TD-2    & CRED & legacy-lab-archive    & TD & \texttt{lab-archive-token}; terminal lab-chain node \\
CRED-TD-3    & CRED & legacy-settlement-api & TD & \texttt{swift-integration-secret} via service-account binding \\
CRED-TD-4    & CRED & eligibility-checker   & TD & \texttt{payer-network-token}; cross-namespace chain \\
CRED-TD-5    & CRED & ops-console          & TD & \texttt{ops-console-credentials}, key \texttt{password} \\
\midrule
NET-TD-1     & NET & api-gateway          & TD & Egress names exact callee labels from call graph \\
NET-TD-2     & NET & prescription-service & TD & Egress to three named callees (patient, lab, pharmacy) \\
NET-TD-3     & NET & notification-service & TD & Ingress must name caller appointment-service \\
NET-TD-4     & NET & auth-service         & TD & Ingress must name caller api-gateway \\
NET-TD-7     & NET & ai-inference         & TD & Egress to imaging-store, radiology-report; deny external \\
\midrule
POLICY-TD-1  & POLICY & claims-processor   & TD & Provenance annotation; image must not change \\
POLICY-TD-2  & POLICY & identity-provider  & TD & Seccomp profile (\texttt{RuntimeDefault} accepted) \\
POLICY-TI-1  & POLICY & session-cache      & TI & Add cost-center annotation (any value) \\
\midrule
RBAC-TD-1    & RBAC & lab-service          & TD & \texttt{resourceNames} must name \texttt{lab-api-key} \\
RBAC-TD-2    & RBAC & billing-service      & TD & Scope to \texttt{billing-rates-config} ConfigMap \\
RBAC-TD-5    & RBAC & identity-provider    & TD & Service-account bound to \texttt{idp-signing-key} Secret \\
RBAC-TD-6    & RBAC & ops-console          & TD & Scope \texttt{pods/exec} to \texttt{audit-logger-dep} \\
RBAC-TI-1    & RBAC & notification-service & TI & Delete named ClusterRole (in finding) \\
RBAC-TI-2    & RBAC & patient-service      & TI & Delete named ClusterRole (in finding) \\
\midrule
SEC-TD-1     & SEC & pharmacy-gateway      & TD & \texttt{ext-pharm-api-secret} via service-account binding \\
SEC-TD-2     & SEC & appointment-service   & TD & \texttt{payment-webhook-secret} via service-account binding \\
SEC-TD-3     & SEC & notification-service  & TD & \texttt{notification-credentials}, key \texttt{password} \\
SEC-TD-4     & SEC & auth-service          & TD & \texttt{auth-credentials}, key \texttt{password} \\
SEC-TD-5     & SEC & fhir-gateway          & TD & \texttt{ehr-integration-token}; clinical namespace \\
\midrule
STORAGE-TD-1 & STORAGE & patient-db        & TD & \texttt{StorageClass} \texttt{phi-storage}, cluster-only \\
STORAGE-TD-2 & STORAGE & imaging-store     & TD & Separate per-service PVCs (dicom, ai) \\
STORAGE-TI-1 & STORAGE & audit-logger-dep  & TI & Add \texttt{emptyDir} size limit (in finding) \\
\bottomrule
\end{tabular}
\end{table}